\documentclass[english,12pt]{article}
          \usepackage{babel}
          \usepackage[T1]{fontenc}
          \usepackage[latin2]{inputenc}
          \usepackage{pslatex}
\begin{document}

\title{PF120916 Piecki fireball and Reszel meteorite fall}

\author{
        A. Olech$^1$, P. \.Zo{\l}\c{a}dek$^2$, Z. Tymi\'nski$^3$, 
M. Stolarz$^2$, M. Wi\'sniewski$^2$, \\
M. B\c{e}ben$^2$, T. Lewandowski$^2$, K. Polak$^2$, A. Raj$^2$
and P. Zar\c{e}ba$^2$}

\date{
          $^1$Nicolaus Copernicus Astronomical Center, 
Bartycka 18, 00--716 Warszawa, Poland, email: olech@camk.edu.pl\\
          $^2$Comets and Meteors Workshop, ul. Bartycka 18, 00-716 Warszawa, Poland\\
          $^3$Narodowe Centrum Bada\'n J\k{a}drowych, O\'srodek Radioizotop\'ow POLATOM, 
ul. So{\l}tana 7, 05-400 Otwock, Poland
          }
\maketitle

\begin{abstract}

On September 12, 2016, at 21:44:07 UT, a $-9.2\pm0.5$ mag fireball appeared
over northeastern Poland. The precise orbit and atmospheric trajectory
of the event is presented, based on the data collected by six video
stations of {\sl Polish Fireball Network (PFN)}. The PF120916 Piecki
fireball entered Earth's atmosphere with the velocity of $16.7\pm0.3$ km/s and
started to shine at height of $81.9 \pm 0.3$ km. Clear deceleration
started after first three seconds of flight, and the terminal velocity
of the meteor was only $5.0\pm0.3$ km/s at height of $26.0 \pm 0.2$ km.
Such a low value of terminal velocity indicates that fragments with
total mass of around $10-15$ kg could survive the atmospheric passage and
cause fall of the meteorites. The predicted area of possible meteorite
impact is computed and it is located south of Reszel city at
Warmian-Masurian region. The impact area was extensively searched by
experienced groups of meteorite hunters but without any success.

\end{abstract}

\section{Introduction}

Multi-station meteor observations have been conducted for several dozen
years. It is a well-tried and proven method for detecting this kind of
phenomena, determining their absolute brightness, meteoroid mass,
trajectory in the atmosphere, orbital parameters and also calculating
the places of meteorite falls. During the last 50 years of 20th century
the European Fireball Network was leading in this area, with their fully
automated photographic stations (Spurn\'y 1997, Spurn\'y et al. 2007).

The {\sl Polish Fireball Network (PFN)} is the project established in
2004, which main goal is to constantly monitor the sky over Poland in
order to detect bright fireballs occurring over the whole territory of
the country (Olech et al. 2006, Wi\'sniewski et al. 2016). It is run by
amateur astronomers associated in {\sl Comets and Meteors Workshop
(CMW)} and coordinated by astronomers from Copernicus Astronomical
Center in Warsaw, Poland. Today there are over 35 fireball stations
belonging to {\sl PFN} that operate during each clear night. In total
over 70 sensitive CCTV cameras with fast and wide angle lenses are used.

In this paper we report an analysis of the multi-station observation of
the PF120916 Piecki fireball made by cameras of the {\sl Polish Fireball
Network}. The trajectory, orbit and possible area of meteorite fall are
calculated.

\section{Observations and data reduction}

The PF120916 Piecki fireball was observed by six {\sl PFN} video
stations. They are listed in Table 1 together with their respective coordinates and
equipment used for recording the fireball. The most detailed recording
comes from the PFN52 Stary Sielc station which works using digital Full
HD ($1920 \times 1200$ pix) camera with recording speed of 25 frames per second. It captured
the meteor almost in the center of the field of view (see Fig. 1). Valuable
recordings were also obtained by PFN47 Jezi\'orko, PFN72 Ko\'zmin
Wielkopolski and PFN74 Brwin\'ow stations which works in PAL resolution
$786\times584$ with 25 frames per second, offering 0.04 second temporal
resolution. Data from PFN48 Rzesz\'ow were not included into the
calculations due to the location of the station along the trajectory of
the fireball, large distance (almost 500 km) from the fireball and
capturing it through the clouds. Location of the PFN63 {\L}\'od\'z
creates substantial other convergence angle with respect to three
stations: PFN52, PFN74, PFN47. In spite of this we did not decide to
use its data for computation, mainly because of low image scale. The
PFN63 is all-sky station equipped with circular fish-eye lens.
Combination of the vertical field of view of 180 deg with
1200 pixels of vertical resolution gives the image scale of around 9 arc
minutes per pixel. Additional reason was large distance (around 300 km)
from the fireball and capturing only part of its trajectory (terminal
point is missing due to the obstruction near the horizon).

\begin{table*}
\small
\centering
\caption[]{Basic data on the PFN stations which observed the PF120916 Piecki fireball}
\vspace{0.1cm}
\begin{tabular}{|l|l|c|c|c|l|l|}
\hline
\hline
Code & Site & Long. & Latit. & Elev. & Camera & Lens \\
 &  & [$^\circ$] & [$^\circ$] & [m] &  & \\
\hline
PFN47 & Jezi\'orko & 21.0572 & 51.9927 & 125 & Siemens CCBB1320-MC & Ernitec 4mm f/1.2 \\
PFN48 & Rzesz\'ow & 21.9220 & 50.0451 & 230 & Tayama C3102-01A1 & Computar 4mm f/1.2 \\
PFN52 & Stary Sielc & 21.2923 & 52.7914 & 90 & DMK23GX236 & Tamron 2.4-6mm f/1.2 \\
PFN63 & {\L}\'od\'z & 19.4795 & 51.6865 & 189 & DMK23GX236 & Lensagon 1.4mm f/1.4 \\
PFN72 & Ko\'zmin W. & 17.4548 & 51.8283 & 139 & Tayama C3102-01A4 & Lenex 4mm f/1.2 \\
PFN74 & Brwin\'ow & 20.7068 & 52.1368 & 110 & Mintron 12V6H-EX & Computar 2.6mm f/1.0 \\
\hline
\hline
\end{tabular}
\end{table*}

The data from four stations used for calculations, after a preliminary
conversion, were additionally reduced astrometricaly by the {\sc UFO
Analyzer} application (SonotaCo 2009). Initially only automatic data
were taken into account but during later processing it became obvious
that significant overexposures, presence of the wake and a possible
fragmentation after the flare caused substantial errors in position of
the  points of the phenomenon calculation. To improve the measurement
precision, the bolide's position was determined with help of {\sc
UFOAnalyzer} astrometric solution using manual centroid measurement.

The trajectory and orbit of fireball were computed using {\sc PyFN}
software (\.Zo{\l}\c{a}dek 2012). {\sc PyFN} is written in Python with
usage of SciPy module and CSPICE library. For the purpose of trajectory
and orbit computation it uses the plane intersection method described by
Ceplecha (1987). Moreover, {\sc PyFN} accepts data in both {\sc MetRec}
(Molau 1999) and {\sc UFOAnalyzer} (SonotaCo 2009) formats and allows
semi-automatic search for double-station meteors.

\newpage

~

\vspace{9.9cm}

\includegraphics{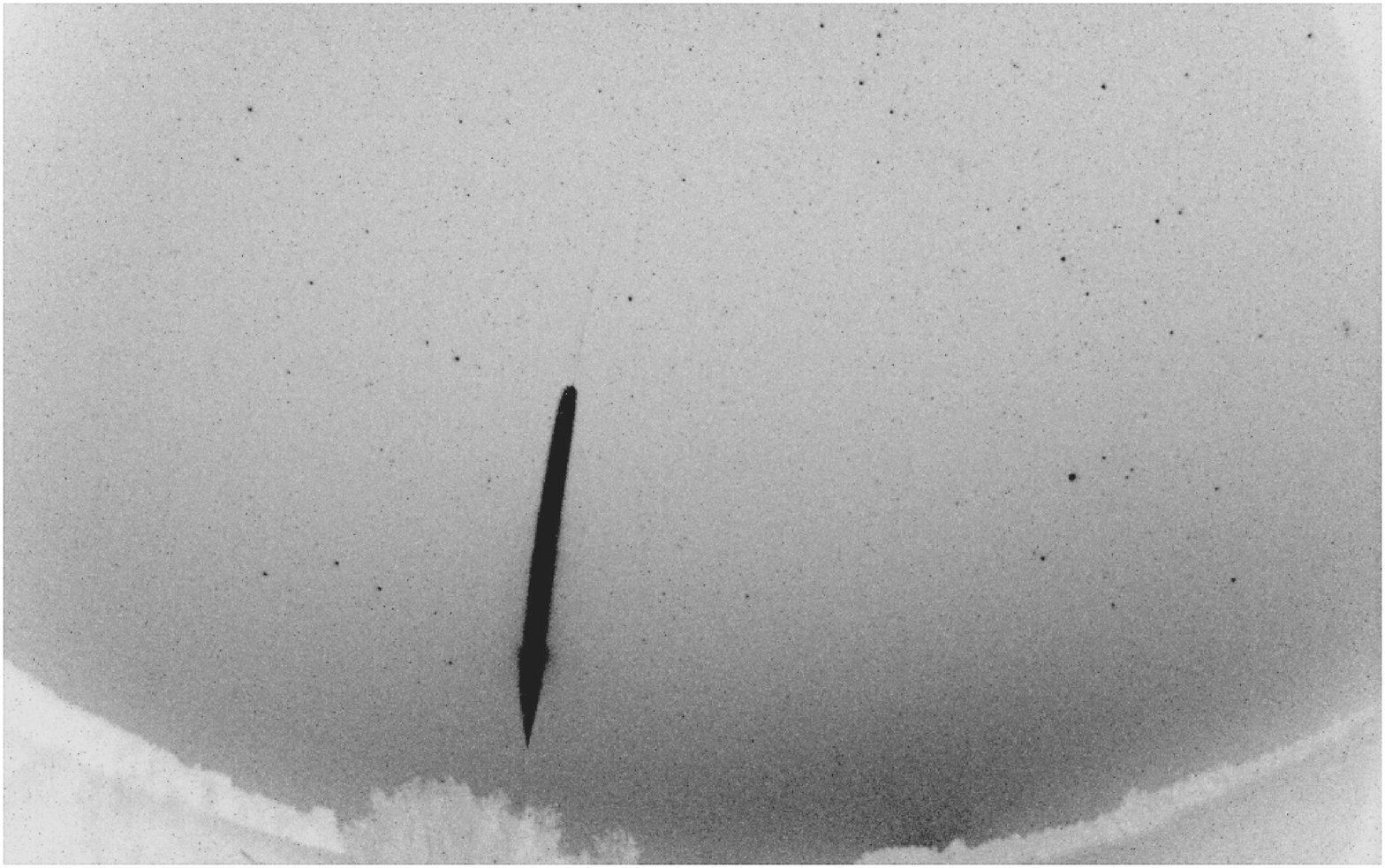}

\begin{figure}[h]
\caption{The video image of PF120916 Piecki fireball captured in
Stary Sielc station (PFN52).}
\end{figure}

\vspace{10cm}

\includegraphics{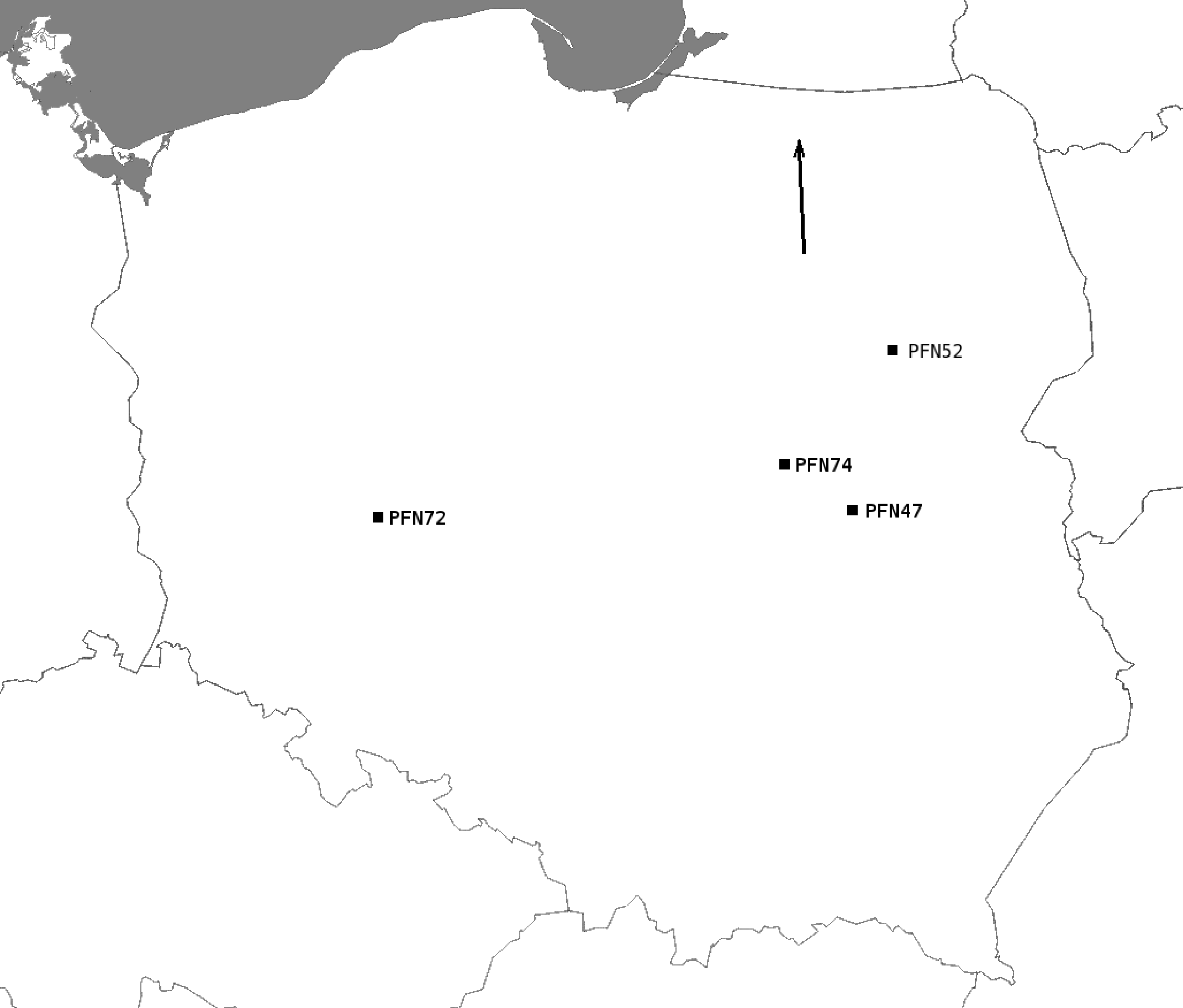}

\begin{figure}[h]
\caption{The luminous trajectory of the PF120916 Piecki fireball over Poland
and the location of the PFN stations which recorded the phenomenon.}
\end{figure}

\section{Results}

\subsection{Trajectory of the fireball}

The Piecki fireball moved almost directly from south to north following
moderately steep trajectory having total length of $91.1\pm1.1$ km. The
beginning of the bolide was located at following
coordinates: $\phi=53.2446(20)^\circ$~N, $\lambda=21.2091(2)^\circ$~E at
the height of $81.9\pm0.3$ km. During next seconds the bolide traveled north
and flew $39.5\pm0.2$ kilometers over Piecki village reaching its
maximum brightness there. The terminal point of the trajectory was
situated at the height of $26.0\pm0.2$ km over the following
coordinates: $\phi=53.884(3)^\circ$~N, $\lambda=21.1504(4)^\circ$~E (10
km west of Biskupiec). The trajectory of the PF120916 fireball is shown
in Fig. 2 and all important parameters are summarized in Table 2.

\begin{table*}
\centering
\caption{Characteristics of the PF120916 Piecki fireball}
\vspace{0.1cm}
\begin{tabular}{lccc}
\hline
\multicolumn{4}{c}{2016 September 12, ${\rm T} = 21^h44^m07^s \pm 1.0^s$ UT}\\
\hline
\multicolumn{4}{c}{Atmospheric trajectory data}\\
\hline
 & {\bf Beginning} & {\bf Max. light} & {\bf Terminal} \\
Vel. [km/s] & $16.7\pm0.3$ & $14.0\pm1.0$ & $5.0\pm0.3$ \\
Height [km] & $81.9\pm0.3$ & $39.5\pm0.2$ & $26.0\pm0.2$ \\
Long. [$^\circ$E] & $21.2091\pm0.0002$ & $21.1638\pm0.0004$ & $21.1504\pm0.0004$\\
Lat. [$^\circ$N] & $53.2446\pm0.002$ & $53.726\pm0.05$ & $53.884\pm0.003$\\
Abs. magn. & $-1.6\pm1.0$ & $-9.2\pm0.5$ & $-4.6\pm1.0$ \\
Slope [$^\circ$] & $38.4\pm0.4$ & $37.9\pm0.4$ & $37.1\pm0.4$\\
Duration & \multicolumn{3}{c}{6.45 sec}\\
Length & \multicolumn{3}{c}{$91.0\pm1.1$ km}\\
Stations & \multicolumn{3}{c}{Stary Sielc, Brwin\'ow, Jezi\'orko, Ko\'zmin Wlk., Rzesz\'ow, {\L}\'od\'z}\\
\hline
\multicolumn{4}{c}{Radiant data (J2000.0)}\\
\hline
 & {\bf Observed} & {\bf Geocentric} & {\bf Heliocentric} \\
RA [$^\circ$] & $341.09\pm0.02$ & $339.83\pm0.06$ & - \\
Decl. [$^\circ$] & $1.66\pm0.36$ & $-6.52\pm0.50$ & - \\
Vel. [km/s] & $16.7\pm0.3$ & $12.5\pm0.3$ & $34.3\pm0.3$\\
\hline
\end{tabular}
\end{table*}

\subsection{Velocity}

Based on our observations the velocity of the phenomena was estimated
for different points of its trajectory. In the initial part of the
flight the velocity did not change in a noticeable way and was almost
constant at the value of around 17 km/s. After the third second of the
flight the velocity decrease became clearly visible with final
deceleration as high as 2457 ${\rm m/s^2}$. At the end of the luminous
trajectory the velocity of the fireball is only $5.0\pm0.3$ km/s (see
Fig. 3).

\newpage

~

\vspace{9.6cm}

\includegraphics{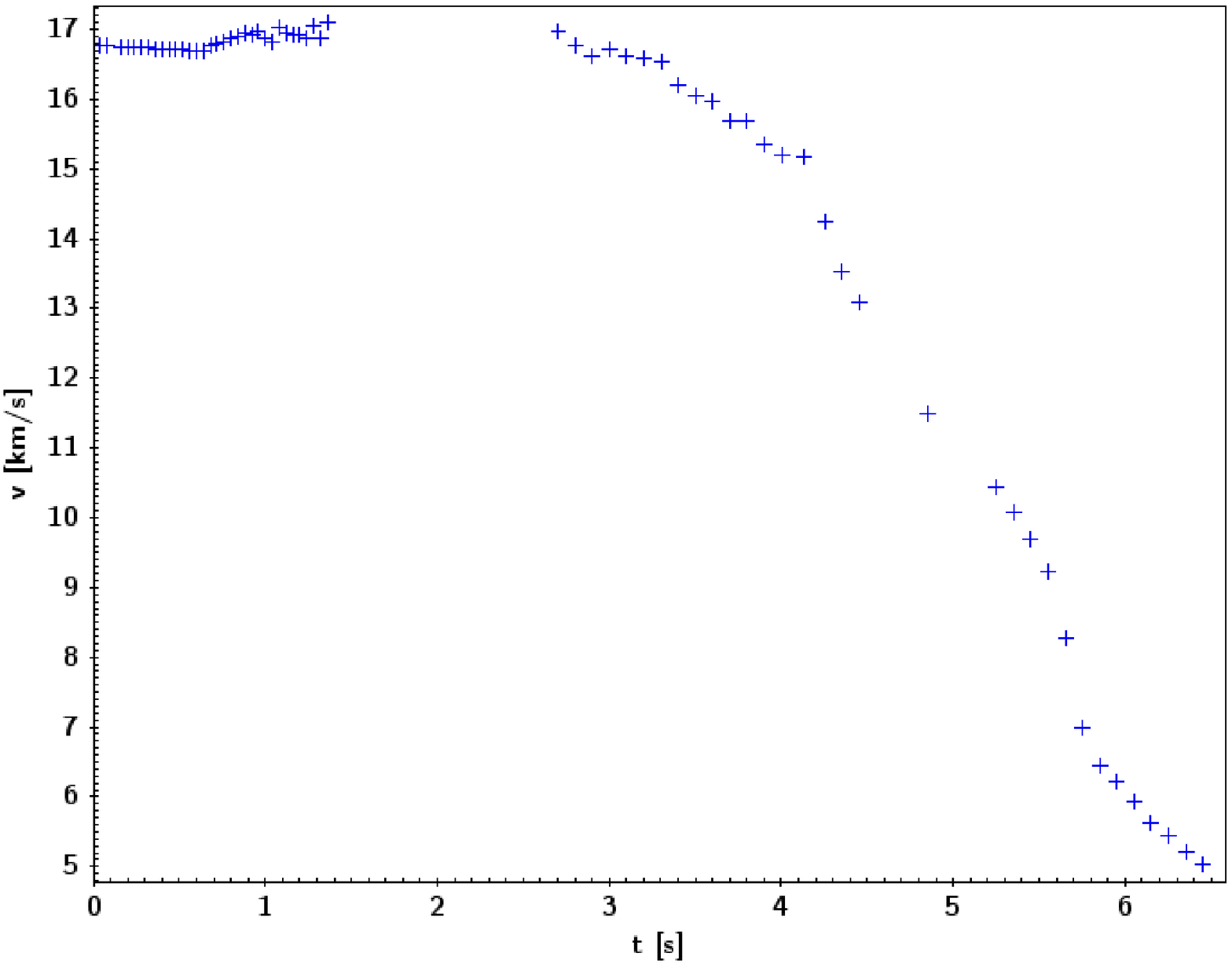}

\begin{figure}[h]
\caption{The evolution of the velocity of the PF120916 Piecki fireball.}
\end{figure}

\vspace{10.7cm}

\includegraphics{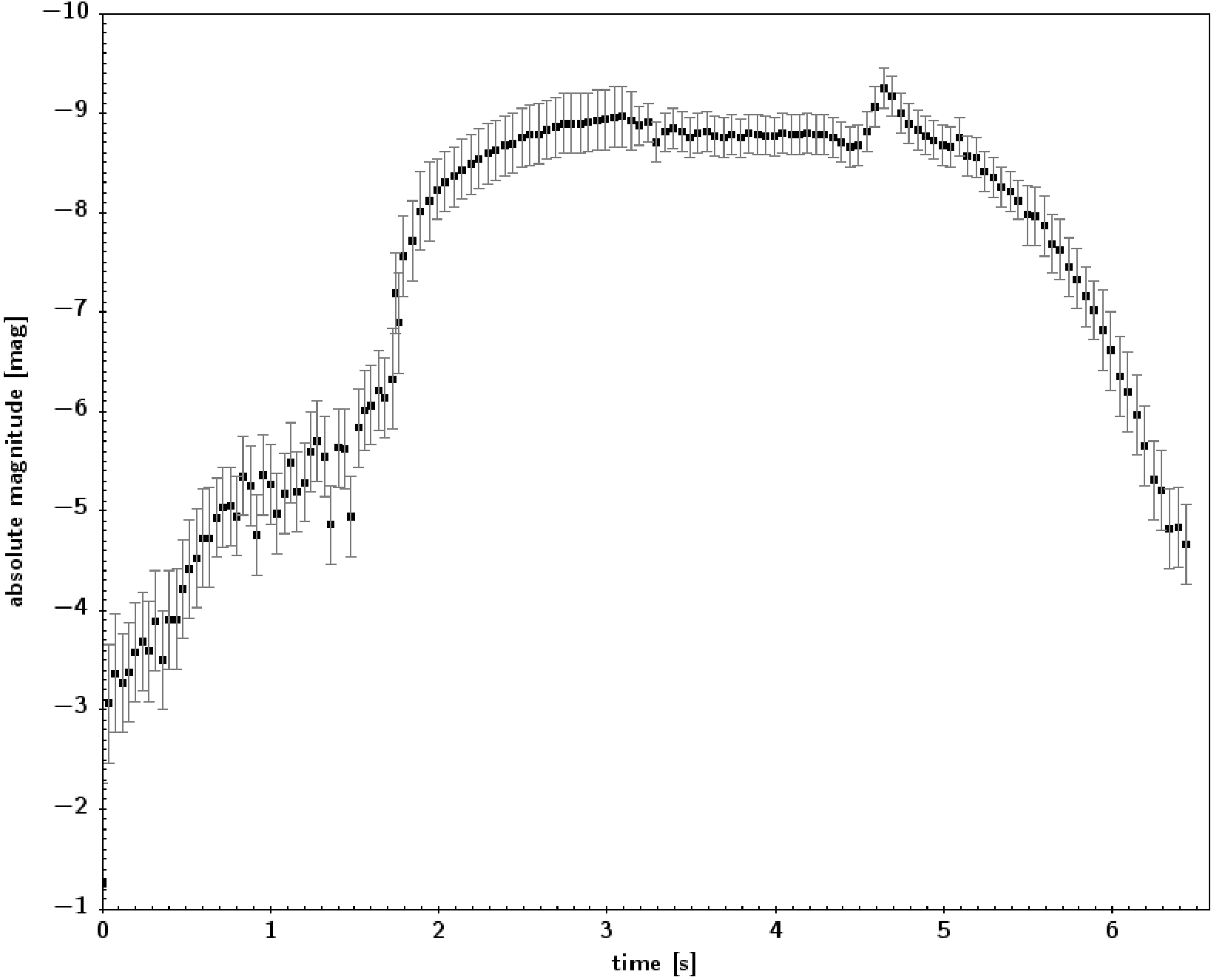}

\begin{figure}[h]
\caption{The light curve of the PF120916 Piecki fireball.}
\end{figure}

\subsection{Brightness}

The photometry of the fireball was not trivial because of the partial
saturation observed. Two independent methods has been applied to
estimate the real brightness of fireball. The first one is based on the
measurements of the sky brightness during the fireball flight. These
measurements have been compared with the sky brigtness caused by the
Full Moon in the same camera. The second method uses for comparison several
objects recorded by cameras of different types with the same optics
working in similar sky conditions. Detailed description of both methods
was given in Olech et al. (2016).

The light curve of the PF120916 Piecki fireball is shown in Fig. 4. At
the first point of the observed trajectory the brightness of the meteor
was estimated to $-1.3\pm1.0$ mag. On the next frame it raised to
$-3.1\pm0.6$ mag. During the first two seconds of flight the
absolute brightness of the meteor increases systematically from $-3.1$
to $-6$ mag. There is a rapid brightening from $-6$ to $-8$ mag around
$t=1.9$ sec. After that moment a plateau lasting about three seconds is
observed with brightness reaching almost $-9$ mag. Maximal absolute
magnitude was detected during small flare at $t=4.62$ seconds with
$m=-9.2 \pm 0.5$.

After appearance of the flare brightness starts to decrease - slowly at
first but then very quickly. At terminal point of the luminous trajectory
the meteor had brightness of $-4.5$ mag.

\vspace{11cm}

\includegraphics{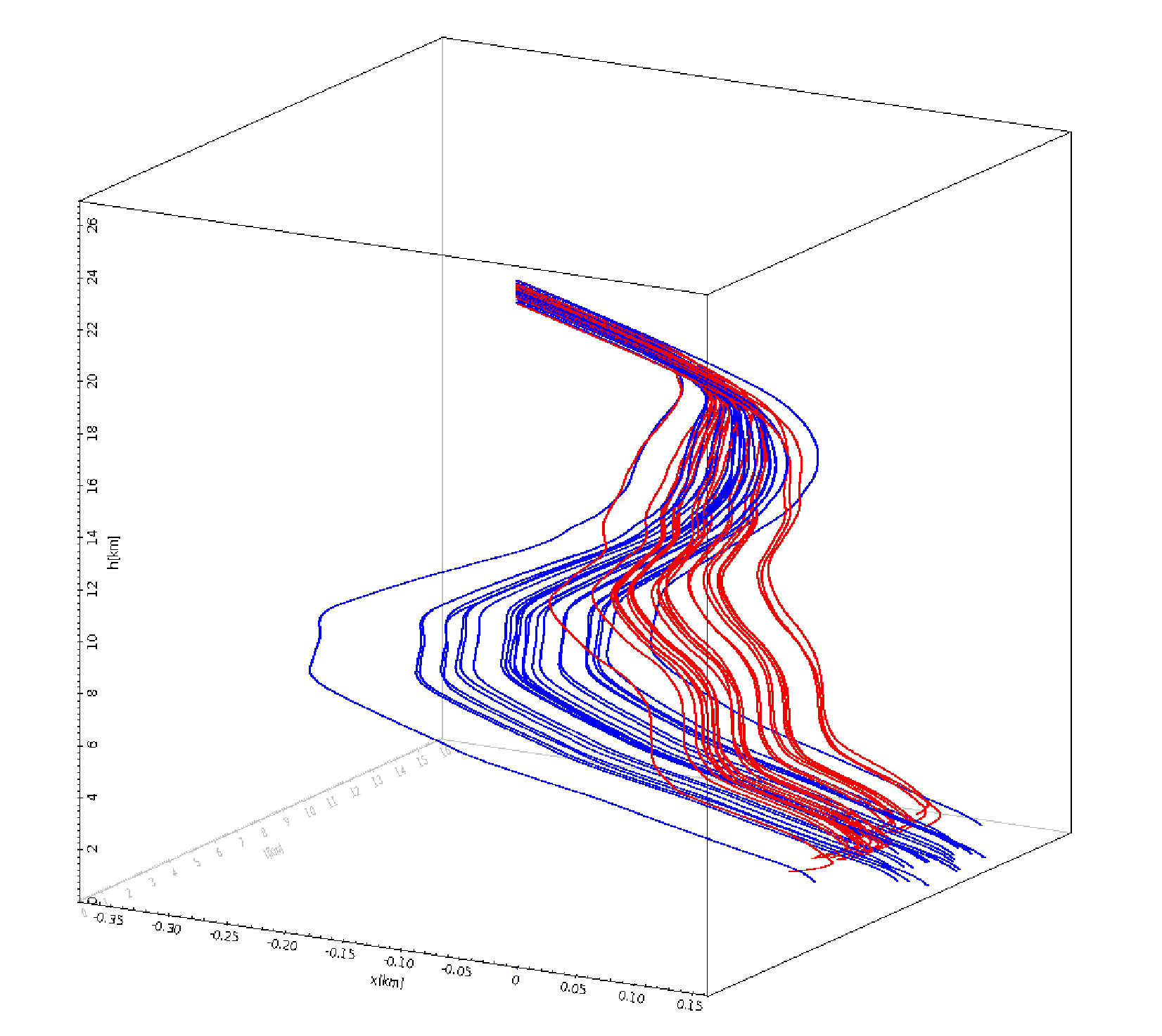}

\begin{figure}[h]
\caption{The dark flight tracks for $2-15$ kg meteorites. 
Different colors of the tracks are used to
distinguish the atmosphere profiles from Kaliningrad and {\L}eba.}
\end{figure}

Both, evolution of the velocity and the light curve of the meteor
indicate that we are dealing with type I of fireball, which is usually
identified with stony material, mostly ordinary chondrities. Taking
into account the occurrence of the fragmentation, the calculated entry
mass is at the level of 50 kg. For this mass, terminal height and entry
velocity the $PE$ parameter (Ceplecha and McCrosky 1976) is around $-4.3$
indicating type I fireball. To include the Piecki fireball to the type
II meteoroids (carbonaceous material with with $-5.25 < PE < -4.6$) one
should increase the initial mass to over 200 kg, which is unlikely in
this case.

\subsection{Dark flight and a possible fall of the meteorite}

The observed final velocity of 5.0 km/s at the height of 26 km indicates
a possible meteorite fall south of Reszel city. The total mass which
survived the atmosphere passage is estimated to be around $10-15$ kg. 
This is dynamical mass estimated from the parameters of the terminal
points of the trajectory of main fragment. We deal here with the most
probable case of meteorite fall in the history of our project.

Calculations for the dark flight in the atmosphere were performed using
the standard method for the European Fireball Network (Ceplecha 1987).
As the initial parameters we used observational data from the final part
of the trajectory such as the terminal velocity, final height,
deceleration, azimuth and inclination. For dark flight calculations
profiles of the atmosphere achieved by balloon from Kaliningrad and
{\L}eba were used. Fortunately, during that night the winds at different
altitudes blew mostly from the north and from the south, thus there were
no large lateral deviations from the original trajectory. 

Fig. 5 shows the computed trajectories for dark flight for meteorites
with mass in range $2-15$ kg. Different colors of the tracks are used to
distinguish the atmosphere profiles from Kaliningrad and {\L}eba.
Despite of significant differences in tracks at height of around 10 km,
the fall points are located in the well defined region.

The calculations were repeated by modifying the input data within the
observed uncertainties. Finally, the potential rectangular area of fall with 
100-200 meters of width and a length of up to four kilometers was
obtained. It is located about 4 km south of Reszel city and couple
hundred meters east of Leginy village overlapping partially with
Legi\'nskie and K{\l}aw\'oj lakes (see Fig. 6).

The light curve of Piecki fireball indicates that meteoroid fragmented 
during the atmosphere passage. Thus, it is highly unlikely that single
15 kg meteorite hit the ground. Most probable impactors have mass of
around 5 kg. 

The bolide event occurred in the northern part of the Warmia and Masuria
region, in the south-western part of the K\c{e}trzyn district in the
mezoregion of  Mr\c{a}gowo Lakes. It is a part of the Mazurian Lake
District with an undulating character of surface, locally hilly with the
inclination towards the north-west. The calculated strewn-field obtained
from the data is situated south of the town of Reszel. The land there 
are meadows and wastelands with two lakes placed in the middle and south
of the area - the K{\l}aw\'oj Lake and Legi\'nskie Lake. The surface
basement consist of a continuous Quaternary layers of clays, silts and
sands lakeside, silts and peats.

The several expeditions to the strewn-field took place in the period of
September to November 2016, involving more than 20 volunteers who
searched the area and walked a total of over 1,200 km. Because the 
meteorite fall was in the late autumn the part of the arable land was
already plowed and the fertile black soil made it difficult to recognize
the meteorite on its surface. Also the most promising area was not
searched enough because of lack of access - the most likely drop of 5 kg
meteorite were in the lush swamp suited south of K{\l}aw\'oj Lake.

One of the search goals was to cover also the region of smallest mass
strewn-field in order to increase the probability of meteorite find.
Unfortunately, the lack of a larger number of searchers, the bad weather
and the diminishing by cultivating still available area did not allow us
for the efficient prospection. The searched out area is marked by tilted
stripes at the Fig. 6.

\newpage

~

\vspace{13.5cm}

\includegraphics{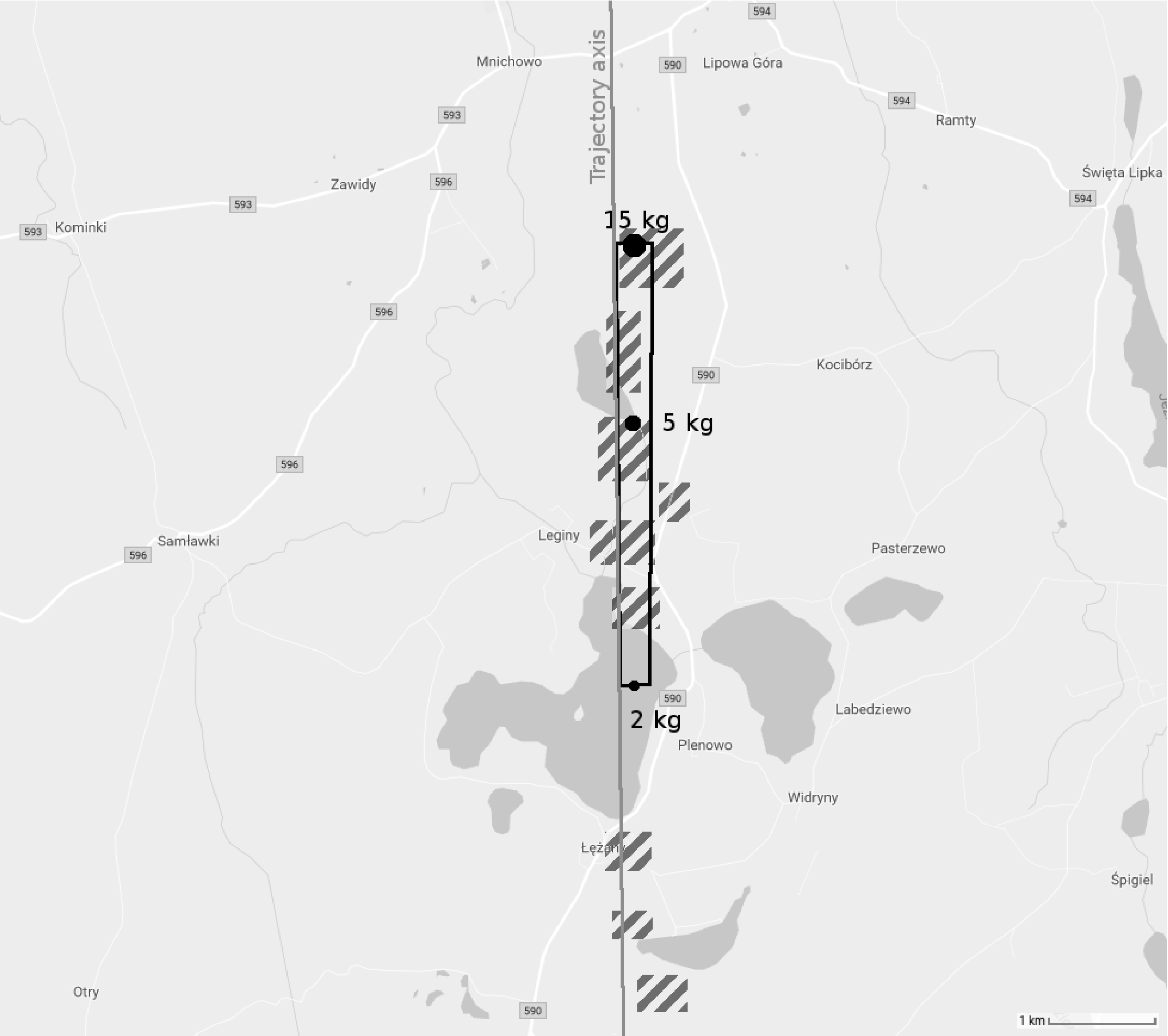}

\begin{figure}[h]
\caption{The computed impact area of Reszel meteorite caused by PF120916 Piecki fireball.
The impact points for 15, 5 and 2 kg meteorites and trajectory axis are also shown.
The area searched by our expeditions is marked by tilted stripes.}
\end{figure}

In the spring time at least one large search campaign is planned by a
group of volunteers. We plan to use metal detector prospection to check
the most probable impact points in the skipped areas during the autumn
campaign. Some volunteers also declared a willingness to individually
scan for meteorites the strewn-field next year.

\subsection{Orbit}

Based on the observational data we were able to determine the radiant of
PF120916 Piecki fireball, its geocentric velocity and orbital 
parameters of the meteoroid which entered the Earth's atmosphere. The
orbital parameters of the meteroid which caused the fireball are listed
in Table 3.

\begin{table*}
\small
\caption[]{Orbital elements of the PF120916 Piecki fireball and asteroids 
with similar orbits ($D'<0.05$). Both the orbital elements of the NEOs and
their errors are taken from the NEODyS-2 database.}
\vspace{0.1cm}
\begin{tabular}{|l|c|c|c|c|c|c|c|}
\hline
\hline
  & $a$ & $e$ & $q$ & $\omega$ & $\Omega$ & $i$ & $D'$ \\
  & [AU] & & [AU] & [deg] & [deg] & [deg] & \\ 
\hline
PF120916   &  1.520(7)  & 0.476(2)   & 0.795(1)  & 249.50(6)  & 170.3150(1) & 0.72(2) &     \\
2014 DA    & 1.5056(4)  & 0.4848(2)  & 0.7757(2) & 110.192(2) & 313.662(1)  & 2.464(1) & 0.0264 \\
2013 BY2   & 1.593(1)   & 0.476(1)   & 0.834(1)  & 298.7(9)   & 114.01(6)   & 2.265(4) & 0.0324 \\
1998 OX4   & 1.58043(1) & 0.48572(1) & 0.8128(1) & 117.114(1) & 299.704(1)  & 4.513(1) & 0.0347 \\
2014 SH224 & 1.5766(1)  & 0.4657(3)  & 0.8424(1) & 345.86(4)  & 80.41(4)    & 0.346(1) & 0.0353 \\ 
2004 RU109 & 1.5325(7)  & 0.4894(3)  & 0.7825(5) & 250.588(2) & 171.395(2)  & 5.849(3) & 0.0355 \\
2013 BR27  & 1.5516(2)  & 0.4594(8)  & 0.8388(2) & 120.72(2)  & 296.97(2)   & 2.565(4) & 0.0373 \\
2014 SS143 & 1.485(1)   & 0.4577(4)  & 0.8054(9) & 259.919(3) & 174.596(1)  & 1.727(2) & 0.0438 \\
2013 CV83  & 1.4346(4)  & 0.4533(2)  & 0.7843(1) & 86.934(8)  & 339.417(7)  & 4.572(2) & 0.0444 \\
2015 RU36  & 1.636(8)   & 0.486(2)   & 0.8412(9) & 75.74(2)   & 349.963(7)  & 4.02(2)  & 0.0446 \\
2013 RN9   & 1.560(3)   & 0.466(1)   & 0.8325(5) & 105.988(9) & 324.51(2)   & 3.50(1)  & 0.0456 \\
\hline
\hline
\end{tabular}
\end{table*}  

Table 3 lists also the orbital elements (taken from the NEODyS-2
database) of asteroids with orbits similar to PF120916 Piecki fireball.
We selected only bodies with Drummond criterion $D'<0.05$ (Drummond
1981). The largest body from this group is (85640) 1998 OX4 which size
is estimated to be 300-600 meters. It is an Apollo group asteroid and a
Mars crosser. It is included in the Minor Planet Center list of
Potentially Hazardous Asteroids (PHAs) as it comes to within 0.05 AU of
Earth periodically.

Similarity of the orbit of Piecki fireball to almost all mentioned
asteroids seems to be rather accidental. Orbits of such size and shape
and with a slight inclination to the ecliptic are characteristic for
whole group of Apollo type asteroids. The only interesting case is the
20-40 meter size 2014 SH224 asteroid, which passed close to the Earth
($\sim 0.1$ AU) on 2016 September 18.94605 UT i.e. only six days after
the appearance of the Piecki fireball, which indicates that both bodies
might be related. Still it is not a decisive argument. To be more
certain one need to do a numerical integration of the orbital parameters
backwards in time in order to test the link between the fireball and
asteroid.
  
\bigskip \noindent {\bf Acknowledgments.}
~This work was supported by the NCN grant number
2013/09/B/ST9/02168.

\end{document}